\documentclass[11pt,a4paper]{article}
\usepackage[T1]{fontenc}
\usepackage[latin1]{inputenc}
\usepackage[english]{babel}
\usepackage{verbatim}   
\usepackage{amsmath}

\usepackage{epsfig}
\usepackage{amssymb}

\newcommand{\be}{\begin{equation}}
\newcommand{\ee}{\end{equation}}

\newcommand{\bea}{\begin{eqnarray}}
\newcommand{\eea}{\end{eqnarray}}
\newcommand{\ba}{\begin{array}}
\newcommand{\ea}{\end{array}}
\newcommand{\beqa}{\begin{eqnarray}}
\newcommand{\eeqa}{\end{eqnarray}}

\newcommand{\lsim}{\lesssim}
\newcommand{\gsim}{\gtrsim}

\newcommand{\matr}{\left( \begin{array}}
\newcommand{\ematr}{\end{array} \right)}

\newcommand{\nue}{{\nu_{e}}}
\newcommand{\num}{{\nu_{\mu}}}
\newcommand{\nut}{{\nu_{\tau}}}

\begin{document}

\mbox{}\vspace*{-1cm}\hspace*{9cm}\makebox[7cm][r]{JyFL-HE 1/2004,
NORDITA-2004-1AP}

\bigskip

\Large

\begin{center}
{\bf Effects of degenerate sterile neutrinos on the supernova
neutrino flux}

\bigskip

\normalsize
{P. Ker\"anen\footnote{keranen@nordita.dk}}

{\it Nordita, Copenhagen, Denmark}

{J. Maalampi\footnote{jukka.maalampi@phys.jyu.fi}}

{ \it Department of Physics, University of Jyv\"askyl\"a, Finland} and
{\it Helsinki Institute of Physics, Helsinki, Finland}

\smallskip
{M. Myyryl\"ainen\footnote{minja.myyrylainen@phys.jyu.fi}} and
{J. Riittinen\footnote{janne.riittinen@phys.jyu.fi}}

{ \it Department of Physics, University of Jyv\"askyl\"a, Finland}
\\[15pt]

\bigskip

\normalsize
{\bf\normalsize \bf Abstract}

\end{center}

\normalsize

We consider the possibility that there exist sterile neutrinos
which are closely degenerate in mass with the active neutrinos and
mixed with them. We investigate the effects of this kind of
active-sterile neutrino mixing on the composition of supernova
neutrino flux at the Earth. If an adiabatic MSW-transition between
active and sterile neutrinos takes place, it could dramatically
diminish the electron neutrino flux.

\bigskip
\newpage

\normalsize
\noindent

\section{Introduction}

Sterile neutrinos closely degenerate with active neutrinos may have, if
they exist, escaped detection in the laboratory and astrophysical
experiments performed so far. They can, however, reveal themselves through
measurable oscillation effects in astronomical-scale
baselines~\cite{Minja03, Pakvasa03}. They may affect the relative fluxes
of the active neutrinos at the Earth if the mass-squared difference obey
$\delta m^2\gsim E/L$, where $L$ is the distance to the source and $E$ is
the energy of neutrinos.  With ultrahigh-energy cosmic ray (UHECR)
neutrinos one can reach the sensitivity of $\delta m^2 \gsim
10^{-18}$~eV$^2$ in the future neutrino telescopes like the
ICECUBE~\cite{Icecube}. It should be emphazised that the mass-squared
differences below $\delta m^2\lsim 10^{-11}$~eV$^2$ cannot be probed in
solar neutrino, atmospheric neutrino or laboratory experiments. The most
stringent present constraint on the active-sterile neutrino mixing comes
from cosmology~\cite{cosmo}, $|\delta m^2|\sin^2 2\varphi <5\times
10^{-8}$~eV$^2$. (These cosmological bounds may be avoided in the case of
large lepton number asymmetries in the early universe \cite{Volkas}). 
It is generally thought that neutrinos are produced in UHECR sources via a
pion-muon decay chain, which yields the flux ratios 
$F^0_e:F^0_\mu:F^0_\tau =1:2:0$, where $F^0_\alpha$ is the the flux of the
neutrino flavour $\nu_{\alpha}$.  If one considers only the active
neutrinos $\nue, \num$ and $\nut$ and takes into account their "bi-large"
mixing behaviour, observed in the solar and atmospheric neutrino
measurements, the flavour ratios of neutrinos at the Earth are predicted
to be $F_e:F_\mu:F_\tau =1:1:1$~\cite{Bento}.  In~\cite{Minja03} we found
that active-sterile mixings with a tiny $\delta m^2$ can change these
ratios at the level of tens of percents. The effects of this size should
be easily detected in the future experiments.

In the present paper we shall extend our analysis of the degenerate
active-sterile mixing to supernova neutrinos. As far as oscillations are
concerned, the situation is for supernova neutrinos quite different from
that for the UHECR neutrinos.  In the case of UHECR neutrinos, the vacuum
oscillations play a central role while for the supernova neutrinos matter
effects are decisive and vacuum oscillations have usually no effects. In
the case of supernova neutrinos the energy is also much lower, so they
could be studied with experiments like HyperK and UNO.

In the dense core of the supernova the neutrino Hamiltonian is extremely
matter dominated. The interaction eigenstates, in which neutrinos are
produced in various weak interaction processes in the core, coincide in a
good accuracy with the eigenstates of this Hamiltonian.  In transit
through the envelope to the surface of the star, the flavour composition
of these eigenstates changes.  Neutrinos also pass through the MSW
resonance regions corresponding to the solar and atmospheric oscillation
scales, which may affect their behaviour. When leaving the star, neutrinos
are in mass eigenstates that consist of different flavours according to
the mixing pattern they have in vacuum, and they will propagate as those
states to the Earth without further oscillations. Consequently, if sterile
neutrinos exist and if they mix with the active neutrinos, they will be
present in the mass eigenstates entering the Earth. They will thereby
affect the fluxes of the active neutrinos measured in neutrino detectors. 

There is an interesting possibility that one or more of the degenerate
active-sterile neutrino pairs encounter a resonant mixing in the outer
skirts of the envelope of the star. This is possible if that pair has
$\vert \delta m^2\vert \gsim 10^{-11}{\rm eV}^2$, which corresponds to the
oscillation length of the order of the giant progenitor star radius. 
Whether or not such a resonant mixing really occurs depends on how well
the adiabaticity conditions are met in the resonance region, which in turn
depends on the details of the density profile of the envelope in its outer
skirts. If it does occur, it may change the flux ratios of neutrinos and
antineutrinos dramatically, in particular if the active-sterile vacuum
mixing angle is small. 

\section{A sterile mixing scenario}

In the three-flavour framework the flavour fields $\nue, \num$, $\nut$ and
the mass eigenfields $\hat\nu_1, \hat\nu_2$, $\hat\nu_3$ are related to
each other through $\nu_l=U_{li}\hat\nu_i$ , where $U$ is a mixing matrix,
in the following parametrized as \be U=\left( \begin{array}{ccc}
  c_{12}c_{13} & s_{12}c_{13} & s_{13} \\
  -s_{12}c_{23}-c_{12}s_{23}s_{13} & c_{12}c_{23}-s_{12}s_{23}s_{13} & s_{23}c_{13} \\
  s_{12}s_{23}-c_{12}c_{23}s_{13} & -c_{12}s_{23}-s_{12}c_{23}s_{13} & c_{23}c_{13} \\
\ea \right), \label{genU} 
\ee 
where $c_{jk} = \cos{\theta_{jk}}$ and
$s_{jk} = \sin{\theta_{jk}}$ (we neglect the possible CP violation).
The solar neutrino data, together with the recent KamLAND reactor
data~\cite{KamLAND}, indicate the mixing angle $\theta_{12}$ to be
bounded into the range $0.50<\theta_{12}< 0.67$~\cite{Holanda}, \cite{hepph0309130}. On
the other hand, observations of atmospheric neutrinos are
consistent with maximal mixing between the mass eigenstates
$\nu_2$ and $\nu_3$, their mixing angle being within the range
$0.64<\theta_{23}< 0.96$ \cite{hepph0309130}, \cite{GonGarz}. The third mixing angle,
$\theta_{13}$, is bounded by the results of CHOOZ~\cite{chooz} and
Palo Verde~\cite{paloverde} to small values, $0\leq\theta_{13}\leq
0.1$.

Let us now assume that there exists three sterile neutrinos $\nu_{s1}$,
$\nu_{s2}$ and $\nu_{s3}$, which mix pairwise with the states $\hat\nu_1$,
$\hat\nu_2$ and $\hat\nu_3$. We denote the new mass eigenstates that
result from this mixing as follows ($i=1,2,3$):  \begin{equation}
\label{mstates} \begin{split}
  \nu_i  & = \cos{\varphi_i}\ \hat{\nu_i} - \sin{\varphi_i}\ \nu_{si}, \\
  \nu'_i & = \sin{\varphi_i}\ \hat{\nu_i} + \cos{\varphi_i}\ \nu_{si}. 
\end{split} \end{equation} The antineutrino states are defined similarly. 
We assume that the mass difference of the states $\nu_i$ and $\nu'_i$ is
so small that in the processes, like particle decays, which are measured
in laboratory experiments, these two states are not distinguished but
appear as a single state with couplings equal to those of the active state
$\hat\nu_i$. There are models where degenerate neutrino pairs may
naturally appear, see e.g.~\cite{vissani}. It should be emphasized that in
some of these models like in the well-known Pseudo-Dirac model, the mixing
angles are naturally close to their maximal value of $\frac{\pi}{4}$.

In the presence of the sterile neutrinos, the neutrino mixing
matrix~\eqref{genU} is modified to the form

\be
U^{(6)} = \left(%
\begin{array}{cccccc}
  \cos{\varphi_1}\, U_{e1} & \cos{\varphi_2}\, U_{e2} & \cos{\varphi_3}\,  U_{e3}
& \sin{\varphi_1}\, U_{e1}  & \sin{\varphi_2}\, U_{e2} &
\sin{\varphi_3}\, U_{e3} \\
  \cos{\varphi_1}\, U_{\mu 1} & \cos{\varphi_2}\, U_{\mu 2} & \cos{\varphi_3}\,  U_{\mu 3} & \sin{\varphi_1}\, U_{\mu 1}  & \sin{\varphi_2}\, U_{\mu 2} & \sin{\varphi_3}\, U_{\mu 3} \\
  \cos{\varphi_1}\, U_{\tau 1} & \cos{\varphi_2}\, U_{\tau 2} &\cos{\varphi_3}\, U_{\tau 3} & \sin{\varphi_1}\, U_{\tau 1} & \sin{\varphi_2}\, U_{\tau 2} & \sin{\varphi_3}\, U_{\tau 3} \\
  -\sin\varphi_1 & 0 & 0 & \cos{\varphi_1} & 0 & 0 \\
  0 & -\sin\varphi_2 & 0 & 0 & \cos{\varphi_2} & 0 \\
  0 & 0 & -\sin\varphi_3 & 0 & 0 & \cos{\varphi_3} \\
  \end{array}%
\right).
\label{U4}
\ee
Obviously, this is just one possible way to realizise the mixing between
three active and three sterile neutrinos, not the most general case.

\section{Fluxes of supernova neutrinos}

We shall now study the effects of the sterile neutrinos on the fluxes of
supernova neutrinos. As mentioned, for supernova neutrinos matter effects
play the main role, in contrast with the case of UHECR neutrinos, where
vacuum oscillations are important. Because of the effects of matter, the
flavour composition of the supernova neutrino flux observed at the Earth
differs from that in the production region (see, e.g.
~\cite{DigheSmirnov00,lunardini}). 

The neutrinos and antineutrinos may undergo matter enhanced MSW
transitions inside the star if $\delta m^2 \lesssim 10^4$~eV$^2$. In the
case of three active neutrinos there are two possible MSW-resonance
regions, at the densities 
\begin{eqnarray} \rho_{\rm H}\sim 10^3 - 10^4\,
{\rm g/cm}^3\, ,\ \rho_{\rm L}\sim 10-30 \, {\rm g/cm}^3 \, .
\label{densitiesHL} 
\end{eqnarray} 
The subscript H refers to the so-called
high-resonance region, which corresponds to the atmospheric neutrino
oscillation ($\delta m_{\rm
atm}^2=2.6\times10^{-3}$~eV$^{2}$,\,$\sin^{2}\theta_{\rm atm}=0.52$
\cite{hepph0309130}), and the subscript L refers to the low-resonance
region, which corresponds to the large mixing angle (LMA)  solar neutrino
oscillations ($\delta m^2_\odot=6.9\times10^{-5}$~eV$^{2}$ and
$\tan^{2}\theta_{\odot}=0.43$ \cite{hepph0309130}). In the case of the
normal mass hierarchy ($m_1\lsim m_2\ll m_3$) both resonances occur for
neutrinos, whereas in the case of the inverse mass hierarchy ($m_3\ll
m_1\lsim m_2$)  the high-resonance occurs for antineutrinos and
low-resonance for neutrinos. If the system is not fully adiabatic, the MSW
effect is not complete but a level crossing from one matter eigenstate to
another will occur in the resonance region. The level crossing
probability, the so-called Landau-Zener probability, is given in
~\cite{LandauZener}.

The matter effects depend on the density profile of the progenitor star.
It can be shown (e.g.~\cite{lunardini}) that propagation through the
low-resonance region, determined by the solar neutrino parameters, is
adiabatic, i.e. the Landau-Zener probability for it is $P_{\rm L}=0$. For
the high-resonance one can distinguish three cases, defined by the values
of $\sin^2 \theta_{13}$ and neutrino energy and differing in the values of
the Landau-Zener probability ($P_H$) ~\cite{lunardini}:  \begin{enumerate}
\item Adiabaticity breaking region: $\sin^2\theta_{13}\lesssim
10^{-6}\times (E/10 \ {\rm MeV})^{2/3}$, where $P_{\rm H}\approx 1$; 
\item Transition region: $\sin^2\theta_{13}\sim (10^{-6}-10^{-4})\times
(E/10 \ {\rm MeV})^{2/3}$, where $0\lesssim P_{\rm H}\lesssim 1$;  \item
Adiabatic region: $\sin^2\theta_{13}\gsim 10^{-4}\times (E/10 \ {\rm
MeV})^{2/3}$, where $P_{\rm H}\approx 0$.  \label{adiabaticity}
\end{enumerate} In the adiabaticity breaking region the H-resonance has no
effect to the evolution of the neutrino states, whereas in the adiabatic
region a full conversion will occur. For simplicity, we will concentrate
in what follows on these two extreme cases 1 and 3, omitting the
transition region case 2.

In the standard case of three active neutrinos with the normal mass
hierarchy the fluxes of the mass eigenstates leaving the star and entering
the Earth later on, which we denote by ${{F}}=(F_1,F_2,F_3)$, are obtained
from the production fluxes of the flavour eigenstate in the supernova
core, denoted by ${F}^0=(F^0_e,F^0_\mu,F^0_\tau)$, through \cite{AkhmedovLuSmi} \be
{F}={P}{F}^0\,, \label{massflux}\ee where the matrix \be
{P}=\begin{pmatrix} P_H P_L & 1-P_L & (1-P_H)P_L\\ P_H(1-P_L) & P_L &
(1-P_H)(1-P_L)\\ (1-P_H) & 0 & P_H \end{pmatrix} \, , \label{Pmatrix} \ee
describes the conversion probability inside the star. The fluxes of
different flavour eigenstate neutrinos $\nu_{\alpha}$ at the Earth are
then given by \be F^{SM}_{\alpha}=\sum_{i=1}^{3}\vert U_{\alpha i}\vert ^2
F_i.  \ee

Let us now consider the situation in the case of three additional sterile
neutrinos. Sterile neutrinos are not produced in the weak interaction
processes taking place in the core of a supernova. Therefore, the fluxes
of the mass eigenstates on the surface of the supernova are, like in the
standard case, determined by the production rates of the active neutrinos
in the core.  Nevertheless, during the propagation of neutrinos through
the envelope of the supernova, sterile components
$\nu_{s1},\nu_{s2},\nu_{s3}$ are developed as a result of the
active-sterile mixing, that is, active-to-sterile transitions take place. 
The flavour composition of the neutrino flux on the surface and at the
Earth is given by \be F_{\alpha}=\sum_{i=1}^{6}\vert U^{(6)}_{\alpha
i}\vert ^2F_i =\sum_{i=1}^{3}\cos^2\varphi_i\vert U_{\alpha i}\vert ^2F_i
+ \sum_{i'=1}^{3}\sin^2\varphi_{i'}\vert U_{\alpha i'}\vert ^2F_{i'} \ee
where $\alpha=e,\mu,\tau,s1,s2,s3$ and $F_i$'s are given in Eq.
{(\ref{massflux})} instead that $P$ is now six-dimensional and
$F^0=(F^0_e,F^0_\mu,F^0_\tau,0,0,0)$. If the active-sterile mixing angles
$\varphi_i$ are all equal, the relative fluxes of different flavours,
$F_{\alpha}$, do not differ from those of the standard
case, $F^{SM}_{\alpha}$. In particular, this is true in the case where all
the mixing angles $\varphi_i$ are close to their maximal value of
$\frac{\pi}{4}$, as predicted by some models.

In the case of the normal mass hierarchy, antineutrinos do not encounter
MSW resonant conversions, and they end up to different mass eigenstates
than the corresponding neutrinos. The counterpart of the matrix $P$ for 
antineutrinos is a unit matrix. Hence the
active-sterile neutrino mixing affects antineutrinos and neutrinos
differently, and consequently the ratio $F_{\alpha}/F_{\overline\alpha}$
generally differs from its value in the nonsterile case.

In the case of the inverted mass hierarchy, the high-resonance is
encountered by antineutrinos and the low-resonance by neutrinos. 
Obviously, the fluxes of neutrino mass eigenstates on the surface of the
star are obtained from Eq. {(\ref{massflux})} by setting $P_H=1$. It is
also straightforward to see that the antineutrino counterpart of the
matrix $P$ given in Eq. (\ref{Pmatrix}), denoted by $\bar P$, is obtained
in the inverted mass hierarchy case from the matrix $P$ by replacing $P_H$
with $\bar P_H$ (actually $P_H=\bar P_H$ \cite{Fogli}) and $P_L$ with
$1-\bar P_L$ \cite{AkhmedovLuSmi}. The fluxes of the active antineutrinos
at the Earth are then given by \be \bar
F_{\alpha}=\sum_{i=1}^{3}\cos^2\varphi_i\vert U_{\alpha i}\vert ^2\bar F_i
+ \sum_{i'=1}^{3}\sin^2\varphi_{i'}\vert U_{\alpha i'}\vert ^2\bar F_{i'}
\ee where $\bar F^0=(\bar F^0_e,\bar F^0_\mu,\bar F^0_\tau,0,0,0)$ are the
fluxes of antineutrino flavours in the production region and $\bar F_i'$s
on the right-hand side are obtained by \be \bar F =\bar P \bar F^0.  \ee
Although the low-resonance point itself does not appear in the
antineutrino sector, a conversion effect is possible at the low-density
region, and $\bar P_L$ takes into account the adiabaticity of this
transition \cite{DigheSmirnov00}. In our calculations we will assume $\bar
P_L$=0. 

\section{Results}

In the following we shall present numerical estimates for the flux ratios
of supernova neutrinos at the Earth in the case of degenerate sterile
neutrinos, and compare them with the corresponding results in the
standard, nonsterile case. We allow the ordinary mixing angles
$\theta_{ij}$ vary in their phenomenologically allowed regions quoted in
Section 2. The active-sterile mixing angles $\varphi_i$ are allowed to
vary arbitrarily in the range from $\varphi_i=0$ (no mixing) to
$\varphi_i=\pi/4$ (maximal mixing). 

There exists a considerable uncertainty concerning the initial fluxes of
neutrino flavours in the production region in the core of the supernova. 
The value of the ratio $F_e^0:F_{\bar{e}}^0:F_x^0$, where
$F_x^0=F_\mu^0=F_{\bar{\mu}}^0=F_\tau^0=F_{\bar{\tau}}^0$, varies
according to the model one uses~\cite{models}. Traditionally these fluxes
are supposed to be equal, i.e. $F_e^0:F_{\bar{e}}^0:F_x^0=1:1:1$, which is
one reference value we will use in our analysis. According to recent
detailed studies of microprocesses taking place in supernova core, the
flux ratios may considerably differ from this simple assumption. From the
results of \cite{raffelt} we obtain, after integrating over the energy
spectrum, the ratios $F_e^0:F_{\bar{e}}^0:F_x^0=4:3:2$, which we will use
in the following. 

Let us first assume that no active-sterile matter conversion takes place
in the outer skirts of the star.  The opposite case will be considered in
the end of this chapter. We compute the flux ratios $F_e/F_a$,
$F_e/F_{\bar{e}}$, and $F_{\bar{e}}/F_a$, where $F_a$ is the sum of the
fluxes of active neutrinos other than $\nu_e$ and $\bar{\nu}_e$. The
results are shown in tables~\ref{table1} and \ref{table2}.  In
table~\ref{table1} we have compared the values of relative fluxes with and
without active-sterile vacuum mixing, for the normal and inverted mass
hierarchy and for an adiabatic and non-adiabatic conversion at the
high-resonance. We see immediately, that in the case of non-adiabatic
transition in the high-resonance the ratios are independent of the mass
hierarchy. This is in agreement with the conclusion of \cite{lunardini},
that for $\sin^2 \theta_{13}<10^{-6}$ the observation of supernova
neutrinos are insensitive to the mass hierarchy.  In adiabatic case, the
relative amount of electon neutrinos is larger for the inverted hierarchy. 
The mixing with sterile neutrinos widens the range of the possible values
of the flux ratios in all cases.  With equal initial fluxes the situation
is more simple, as shown in table \ref{table2}, in the sense that the
ratios are
independent of the neutrino mass hierarchy and the adiabaticity of the MSW
conversions. Also in this case the active-sterile mixings considerably
widen the range of the possible flux ratios. 

\begin{table}[ht!]
\centering
\begin{tabular}{|c|l|c|c|c|c|}
\multicolumn{6}{c}{} \\ \cline{3-6}

\multicolumn{2}{c}{} & \multicolumn{2}{|c|}{Normal hierarchy} &
\multicolumn{2}{|c|}{Inverted hierarchy} \\ \cline{3-6}
\multicolumn{2}{c|}{} & active & active + sterile & active &
active + sterile\\
\hline
&$\frac{F_e}{F_a}$ & $0.18$ & $0.12-0.28$ & $0.22-0.25$ & $0.16-0.33$\\
adiab.&$\frac{F_{\bar{e}}}{F_a}$ & $0.28-0.29$ & $0.18-0.46$
&$0.20$&$0.14-0.31$\\
&$\frac{F_e}{F_{\bar{e}}}$ & $0.63-0.67$ & $0.61-0.72$ & $1.07-1.21$&
$0.99-1.35$\\ \hline

&$\frac{F_e}{F_a}$ & $0.24-0.27$ & $0.17-0.33$ &0.24-0.27 &0.17-0.33\\
non-adiab.&$\frac{F_{\bar{e}}}{F_a}$ & $0.30$ & $0.20-0.47$ &$0.30$ &
$0.20-0.47$\\
&$\frac{F_e}{F_{\bar{e}}}$ & $0.77-0.92$ & $0.69-1.10$ & $0.77-0.92$
&$0.69-1.10$\\ \hline

\end{tabular}

\caption{Results with initial flux ratios $4:3:2$}
\label{table1}
\end{table}

In a summary, the effect of the active-sterile mixing is in general less
striking for the supernova neutrinos than what we found in our previous
study \cite{Minja03} it to be for the UHECR neutrinos. The relative flux of the
electron antineutrino can, however, differ from its nonsterile value as
much as 50 \% if the active-sterile mixing angles $\varphi_i$ for
different degenerate pairs differ suitably from each other. The basic
reason for this is that the electron neutrino and antineutrino end up to
different mass states in the surface of the star, so that their fluxes are
sensitive to different active-sterile mixing angles. 

\begin{table}[ht!]
\centering
\begin{tabular}{|c|c|c|}
\multicolumn{3}{c}{} \\  \cline{2-3}
\multicolumn{1}{c|}{} & active & active + sterile \\ \hline
$\frac{F_e}{F_a}$ & $0.25$ & $0.17-0.38$ \\
$\frac{F_{\bar{e}}}{F_a}$ & $0.25$ & $0.17-0.38$ \\
$\frac{F_e}{F_{\bar{e}}}$ & $1.00$ & $1.00$ \\ \hline
\end{tabular}
\caption{Results with initial flux ratios $1:1:1$. These values are valid
for both normal and inverted mass hierarchy and independently of
adiabaticity.}
\label{table2}
\end{table}

So far we have assumed that the MSW conversions between the active and
sterile states do not occur, in other words, the transitions between the
mass states $\nu_i = \cos{\varphi_i}\ \hat{\nu_i} - \sin{\varphi_i}\
\nu_{si}$ and $ \nu'_i = \sin{\varphi_i}\ \hat{\nu_i} + \cos{\varphi_i}\
\nu_{si}$ are non-adiabatic. The adiabaticity depends on the profile of
the envelope at the resonance region, as well as on the mixing angles
$\varphi_i$ and the mass-squared difference $m_i^2-m _i'^2$. In order to
get a feeling of what may happen, we have studied, utilizing the analysis of 
ref. \cite{Kachelriess},  a case of mixing of one degenerate active-sterile 
neutrino pair more closely, and we have found that for $m_i^2-m_i'^2\lsim
10^{-11}{\rm eV}^2$ the transition is highly non-adiabatic for mixing angles 
up to about $35^0$ and adiabatic only in the case of nearly maximal mixing. 
For larger mass-squared differencies adiabaticity can be reached with smaller 
mixing angles. In general there will be a range in the parameter values, where 
the transiton is partially adiabatic. Obviously, our analysis does not make 
full justice to the problem, but it does convince us that the sterile mixing of 
the sort we have been looking at may have detectable effects.

Let us now assume that these transitions are fully adiabatic and the
transitions $\nu_i \to \nu'_i $ do take place. In the case the mixing
angle $\varphi_i$ is close to maximal this would not have any influence on
the fluxes of the active neutrinos. In contrast, if the mixing angle
$\varphi_i$ is small, the transition would mean a conversion of a
predominantly active neutrino state into a predominantly sterile state,
which would diminish the observable flux of supernova neutrinos at the
Earth. The mixing angle $\varphi_i$ cannot be, however, arbitrarily small,
since for very small angles the transition would not be adiabatic (unless
the mass-squared difference is large). Large effects are anyway possible,
and if the mixing angles of different degenerate pairs differ from each
other, the flux ratios may change dramatically. This can be seen in Table
3, where the flux ratios are presented for the case of the normal mass
hierarchy and assuming adiabatic $\nu_i \to \nu'_i $ transitions. The
mixing angle $\varphi_i$ is allowed to vary in the range $5^0-45^0$. The
relative flux of the electron neutrino can substatially decrease as a
result of these conversions.

\begin{table}[ht!]
\centering
\begin{tabular}{|c|l|c|c|}
\multicolumn{4}{c}{} \\ \cline{3-4}

\multicolumn{2}{c}{} & \multicolumn{2}{|c|}{Normal hierarchy} \\
\cline{3-4}
\multicolumn{2}{c|}{} & $F^0_e:F^0_{\bar{e}}:F^0_x=4:3:2$ &
$F^0_e:F^0_{\bar{e}}:F^0_x=1:1:1$ \\
\hline
&$\frac{F_e}{F_a}$ & $0.004-0.21$ & $0.005-0.24$\\
adiab.&$\frac{F_{\bar{e}}}{F_a}$ & $0.30-0.64$ & $0.26-0.49$\\
&$\frac{F_e}{F_{\bar{e}}}$ & $0.006-0.64$ & $0.009-0.97$\\ \hline

&$\frac{F_e}{F_a}$ & $0.005-0.25$ & $0.004-0.24$\\
non-adiab.&$\frac{F_{\bar{e}}}{F_a}$ & $0.30-0.66$ & $0.26-0.49$ \\
&$\frac{F_e}{F_{\bar{e}}}$ & $0.009-0.88$ & $0.008-0.97$ \\ \hline

\end{tabular}

\caption{Results with adiabatic $\nu_i \to \nu'_i $ -transitions}
\label{table3}
\end{table}

In considering the significance of the effects discussed above one should
also consider the anticipated accuracy of the experimental determination
of the supernova neutrino fluxes. For example, in the SNO detector, for a
typical supernova in the Galaxy with a distance of 10 kpc, the relative
uncertainty of the $\bar\nu_e$ and $\nu_a$ fluxies is about 5~\% and that
of the $\nu_e$ flux about 10~\% \cite{Vogel}. Consequently, the
uncertainties of the determination of the flux ratios are roughly 5~\% for
$F_{\bar e}/F_a$ and roughly 10~\%-15~\% for $F_{e}/F_a$ and
$F_{e}/F_{\bar e}$. In the Super-Kamiokande detector the $\bar\nu_e$ flux
can be determined more accurately, with the uncertainty of about 1~\%
(corresponding to about observed 10 000 events). The accuracy of the
$\nu_a$ flux determination is estimated to be about 4~\% (about 700
events) \cite{Vogel}. The $\nu_e$ flux is, in contrast, quite hard to
determine at the Super-K. While noting that the effects of the
active-sterile mixing we have discussed would be detectable in these
detectors, we must stress the importance of developing methods of an
accurate determination of the flux of electron neutrinos in future
neutrino experiments. 

\section{Summary}

We have investigated how supernova neutrino fluxes are affected by the
existence of sterile neutrinos closely degenerate with active neutrinos. 
Sterile neutrinos are not produced in the core of a supernova, but in the
matter eigenstates entering the surface of the supernova envelope sterile
components are developed. We have used two sets of initial fluxes of
neutrinos in the production region, $F_e^0:F_{\bar{e}}^0:F_x^0=1:1:1$, and
$F_e^0:F_{\bar{e}}^0:F_x^0=4:3:2$.  Matter effects were taken into account
and we have explored both normal and inverted mass hierarchies. The
effects caused by the degenerate sterile neutrinos will be detectable in
future neutrino experiments. Particularly large effects are possible for
the electron antineutrino flux.

It is possible, at least in principle, that the degenerate mass eigenstate
pairs encounter a MSW resonance conversion in the outer skirts of the
supernova envelope. This may dramatically change the relative fluxes of
the neutrinos interacting in detectors. If the active-sterile mixing
angles is small, a transition from a predominantly active state to a
predominantly sterile state will occur, which can diminish e.g. the
electron neutrino flux a lot. 

\section{Acknowledgements} 
One of us (PK) is grateful to the Department of Physics, University of Jyv\"askyl\"a for
hospitality, and another one (JR) acknowledges the hospitality of NORDITA. 
This work has been supported by the
Academy of Finland under the contract no.\ 40677.

\end{document}